\documentclass[aps,twocolumn,prl,floatfix,longbibliography]{revtex4-1}
\usepackage{amsmath}
\usepackage{amssymb}
\usepackage{bbm}
\usepackage[colorlinks=true,citecolor=blue, linkcolor=blue, urlcolor=blue]{hyperref}
\usepackage{graphicx,color,amssymb}
\usepackage{hyperref}

\begin{document}

\title{Many-body Quantum Chaos and Entanglement in a Quantum Ratchet}

\author{Marc Andrew Valdez$^1$}
\author{Gavriil Shchedrin$^1$}
\author{Martin Heimsoth$^{1,2}$}
\author{Charles E. Creffield$^2$}
\author{Fernando Sols$^2$}
\author{Lincoln D. Carr$^1$}
\affiliation{$^1$Department of Physics, Colorado School of Mines, Golden, CO 80401, USA \\ $^2$Departamento de Fisica de Materiales, Universidad Complutense de Madrid, E-28040, Madrid, Spain}
\date{\today}

\begin{abstract}
We uncover signatures of quantum chaos in the many-body dynamics of a Bose-Einstein condensate-based quantum ratchet in a toroidal trap. We propose measures including entanglement, condensate depletion, and spreading over a fixed basis in many-body Hilbert space which quantitatively identify the region in which quantum chaotic many-body dynamics occurs, where random matrix theory is limited or inaccessible. With these tools we show that many-body quantum chaos is neither highly entangled nor delocalized in the Hilbert space, contrary to conventionally expected signatures of quantum chaos.
\end{abstract}

\maketitle

The ability to engineer and control quantum many-body systems has led to a surge in important advances~\cite{Bloch2008} ranging from quantum computation, to the observation of emergent phenomena and far from equilibrium dynamics~\cite{Sadler2006, Eisert2015}. Specifically, lattice-based Bose-Einstein Condensates (BECs)~\cite{Lignier2007} and spin systems~\cite{Barmettler2009} allow for a deeper grasp of the role interactions and correlations~\cite{Islam2015} play in both static and dynamic phenomena. In the far-from-equilibrium case, these platforms allow exploration of regimes where correlations and fluctuations may be dominant. Moreover, they yield signatures of quantum chaos, which provides insight into the transition from integrable to non-integrable dynamics as well as the quantum-classical correspondence for such dynamics~\cite{Haake2010}. Such a transition to chaos is often associated with quantum chaotic level statistics e.g. the kicked rotor. In the many-body case, a link between quantum chaos and high entanglement has been proposed previously~\cite{Neill2016}, yet the many-body perspective has not been completely investigated. In this Letter, we present dynamical quantum many-body measures including depletion, basis occupation, and entanglement that correlate with quantum chaotic level statistics and provide a general method to identify and quantify quantum many-body chaos in systems where random matrix theory (RMT) is of limited use or inaccessible. We also find that quantum many-body chaos does not require or produce high entanglement, contrary to conventional thinking.

\par
\begin{figure}
\centering
\includegraphics[width=\linewidth]{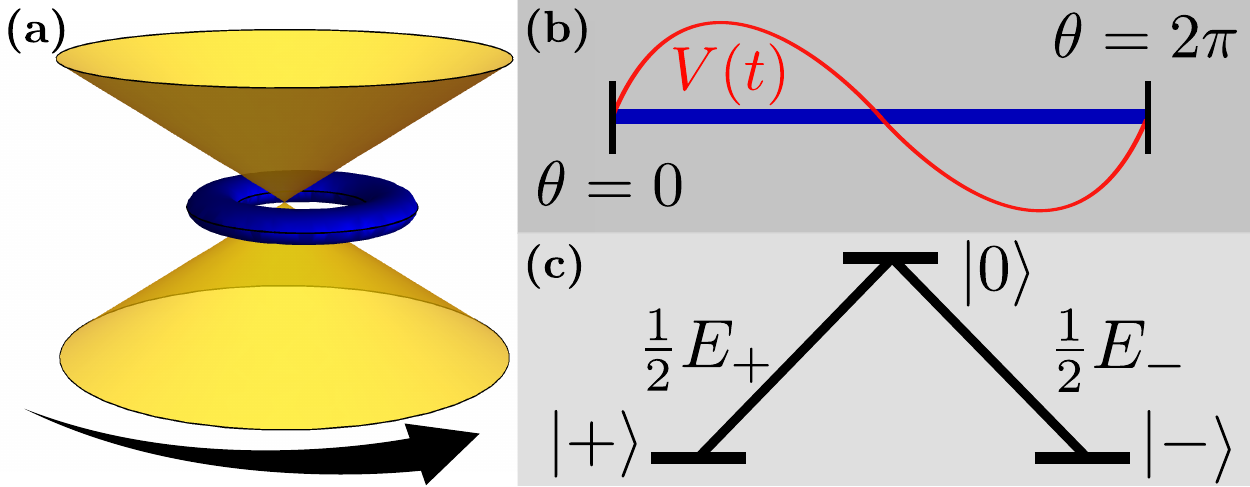}
\caption{\label{scheme}\emph{Quantum Ratchet in a Ring Trap}. (a) BEC (blue torus) trapped in a ring geometry (yellow), being rotated off center (black arrow). The driving breaks generalized P and T symmetries, giving rise to the ratchet effect. (b) The system can be treated as a one-dimensional condensate, driven by the symmetry breaking potential, $V(t)$, generated by the rotation of the trap. (c) The effective three level system, derived from the Floquet-inspired $(t,t')$-formalism~\cite{Heimsoth2012}, provides a simplified treatment of the quantum ratchet. The system exhibits three dynamical regimes in particle current, Rabi oscillations, chaos, and self-trapping, for increasing particle interactions.}
\end{figure}

 As a case study in the properties of quantum many-body chaos, we explore the statics and dynamics of quantum ratchet. This ratchet can be realized as a BEC in a toroidal trap driven by a potential that breaks generalized parity and time-reversal symmetries (Fig.~\ref{scheme}). The violation of these symmetries are well known to produce ratchet effects in the semi-classical limit, i.e., directional motion in the presence of zero time average force, which can be regular or chaotic~\cite{Flach2000, Reimann2002,Denisov2014}. In our system, these regimes are tuned by interaction strength~\cite{Heimsoth2013}. We perform a comprehensive study of the quantum many-body measures of this system in three representations: position, momentum, and a truncated Floquet picture. In this analysis we use entanglement, condensate depletion, and the spreading over a fixed basis in the many-body Hilbert space to quantitatively identify the interaction regime over which quantum chaos begins, ends, and is maximal. Moreover, we show that the RMT level statistics confirm this classification of chaos, although the results rapidly become misleading as more single particle modes are included.

\par Systems ranging from BECs~\cite{Salger2009}, to confined electrons in semiconductors~\cite{Linke1999} and driven graphene~\cite{Drexler2013} can possess non-equilibrium transport properties and have shaped the understanding of the classical and quantum effects leading to such phenomena. Displaying both regular and chaotic dynamics, as well as collective properties~\cite{Cubero2010, Reimann2002}, many-body quantum ratchets provide insight into the role of quantum many-body effects in the transition between regular and chaotic dynamics. Probes into quantum chaos often focus on RMT~\cite{Simons1993,Ponomarenko2008,Gubin2012}, whether in few particle systems~\cite{Zhai2012, Madhok2015} or the many-body case~\cite{Kolovsky2004, Jacquod1997, Mulansky2009}. However, reaching the statistics necessary to classify quantum chaos in such a way frequently becomes intractable due to large amounts of degeneracy or too many dynamically irrelevant single particle modes. Here we provide an in-depth view on the interplay between quantum chaos defined by RMT and entanglement and other measures that are more relevant to interacting many-body systems. Under time evolution these measures provide new methods for which quantum chaos may be identified when RMT analysis becomes out of reach.

\par We consider a BEC in an optical ring trap that is rotated off center, closely following the experimental setups for generating toroidal BECs~\cite{Heathcote2008, Ramanathan2011, Moulder2012, Marti2015}. The rotation of the BEC generates, via inertial effects, a space- and time-dependent potential that breaks the generalized parity and time-reversal symmetries (Fig.~\ref{scheme})~\cite{Heimsoth2013}, generating a ratchet current~\cite{Flach2000, Reimann2002}. In the semi-classical limit it exhibits a continuous transition from Rabi oscillations, to chaos, into self-trapping with the increase of the particle coupling~\cite{Heimsoth2013}. As a tractable quantum many-body generalization of this effectively one-dimensional system, we use a periodic Bose-Hubbard model with fixed particle number. Here the model can be regarded as a discretization of the ring geometry or a ring lattice formed by a toroidal trap in superposition with radial barriers~\cite{Cataldo2011}. Given in hopping units, where the time scale becomes the hopping time $\hbar/J$ and energies are scaled by $J$, we have
\begin{eqnarray}
\hat{H}_\text{B}^\text{Pos}=&-&\textstyle\sum_{i=1}^L( \hat{b}_i^{\dag}\hat{b}_{i+1}+\text{h.c.})+\frac{U}{2}\sum_{i=1}^L \hat{n}_i(\hat{n_i}-\hat{\mathbbm{1}})\nonumber\\
&+&\textstyle\sum_{i=1}^L V_i\hat{n}_i \label{Bose-Hubbard}
\end{eqnarray}
with the driving potential
\begin{equation}
V_i=V_i(t)=E_+\cos(\kappa r \theta_i-\omega t)+E_-\cos(-\kappa r \theta_i-\omega t).
\end{equation}
Here $\hat{b}_{i}$, $\hat{b}_{i}^{\dag}$, and $\hat{n}_i$ are the bosonic annihilation, creation, and number operators, respectively, with periodic boundary conditions $\hat{b}_{L+i}=\hat{b}_i$. $\kappa$ is the wavenumber of the driving field, $\omega$ is the driving frequency, $r$ is the radius of the condensate, $E_\pm$ are the field amplitudes, and $\theta_i\in[0,2\pi)$ is the angle of the $i$th site in an $L$-site discretization. We study resonant driving, $\kappa=1/r$, $\omega=2[1-\cos(2\pi/L)]$. In terms of the mean field interaction strength, $g=2\pi U(N-1)/L$, the transition into self-trapping is given by $g_\text{ST}=4\pi \text{max}(E_+,E_-)$~\cite{Heimsoth2013}. Semi-classically, positive Lyapunov exponents arise in distinct regions on the interval $0.08\leq g \leq g_\text{ST}$, similar to the logistic map having regions of chaos and stability. The time scales for system dynamics are the drive period $T=2\pi/\omega$, and the Rabi period with zero interactions $T_R=2\pi/(E_+^2+E_-^2)^{1/2}$.

\par Without loss of generality we can choose $E_+=9/400$ and $E_-=3/400$, and, unless otherwise specified, $L=10$ with the number of particles $N=5$. When considering dynamics, time evolution calculations are performed using exact diagonalization, with a time step of $0.1T$, and Time Evolving Block Decimation (TEBD)~\cite{Vidal2003, Vidal2004, Schollwock2011,Wall2012,TEBD} with a time step of 0.1$\hbar/J$. The time steps were selected with the convergence of each respective method as the determining factor.

\par
\begin{figure}
\centering
\includegraphics[width=\linewidth]{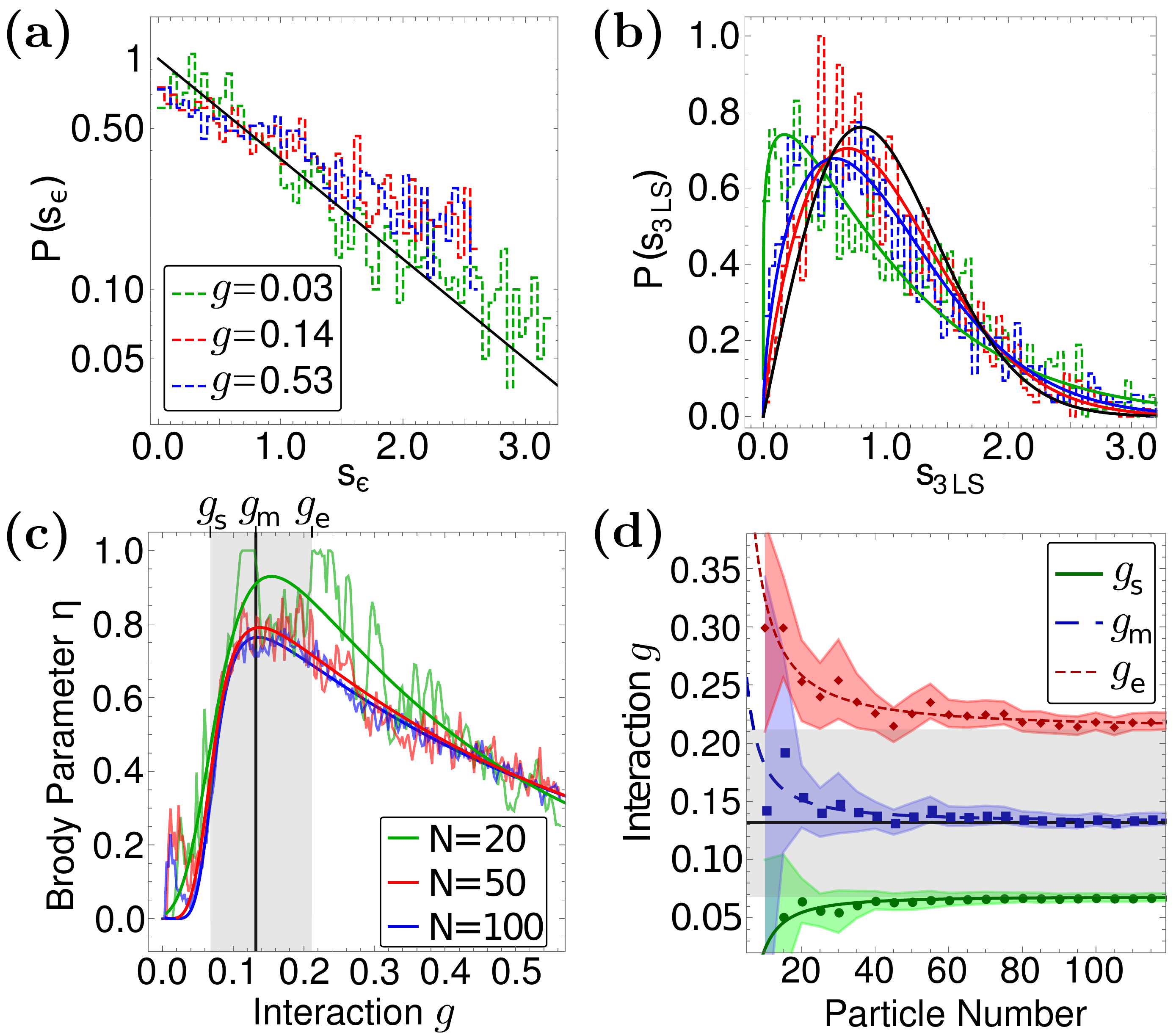}
\caption{\label{levelspacings}\emph{Random Matrix Theory Analysis}. (a) The quasi-energy level statistics of the driven Bose-Hubbard model obtained from the Floquet operator. The dynamical regimes are Rabi oscillations, chaos, and self-trapping for interaction strengths of $g=$ 0.03, 0.14, and 0.53, respectively, and the black line indicates Poisson statistics. The quasi-energy level spacings have increasing probability as the spacing tends to zero for both the regular and chaotic regimes, contrary to the prediction of Random Matrix Theory. (b) Level repulsion is recovered in the truncated three level system, shown for particle number $N=50$. (c) The Brody distribution (solid curves in (b)), which interpolates between Poisson statistics ($\eta=0$) and Wigner-Dyson statistics ($\eta=1$, black curve), has turning points which identify the chaotic regime. The fits ignore the initial fluctuations for $g<0.05$, since they correspond to Rabi dynamics. (d) Critical $g$ as a function of $N$, which asymptotically define the chaotic regime (gray shading in (c)), where the colored regions give the error in $g$.}
\end{figure}

 Since the semi-classical counterpart of our system is chaotic for particular interaction strengths and time periodic, we expect to resolve quantum chaotic level statistics of the Floquet operator $\hat{F}=\hat{U}(T+t_0)\hat{U}(T+t_0-\delta t)\cdots \hat{U}(t_0+\delta t)$, with $\hat{U}(t_i)=\exp[(-i/\hbar) \hat{H}(t_i)\delta t]$~\cite{Haake2010}. RMT predicts that the quasi-energy level spacings $s_\varepsilon$, normalized by the mean spacing, should exhibit level repulsion leading to a probability distribution $P(s_\varepsilon)$ which approach zero polynomially for $s\rightarrow0$. As can be seen in Fig.~\ref{levelspacings}(a), where we have set $\delta t=0.1T$ which well captures $\hat{F}$, the level statistics follow a Poisson distribution that is indicative of regular dynamics. The argument that the near-degeneracies in the quasi-energies wash out the polynomial decrease to zero can be made, yet analysis accounting for near-degeneracies by truncating the lowest level spacings from the calculation revealed no level repulsion.

\par Instead of considering the time-dependent Bose-Hubbard model, we apply the $(t,t')$-formalism~\cite{Peskin1993} to the equation of motion for the field operator $\hat{\psi}(x,t)$ and expand it in the nonlinear Floquet states. Thus, we arrive at a time-independent description of the quantum ratchet~\cite{Heimsoth2012}. Furthermore, previous studies show that three Floquet modes with $k\in\{0,\pm1\}$ capture the dynamics of our quantum ratchet in an effective three-level system~\cite{Heimsoth2013}. This Hamiltonian is then given as
\begin{eqnarray}
\nonumber \hat{H}_{\text{3LS}}=&&\textstyle\frac{1}{2}E_+( \hat{a}^{\dag}_+\hat{a}_0+\text{h.c.})+\frac{1}{2}E_-( \hat{a}^{\dag}_-\hat{a}_0+\text{h.c.})\\&&-\textstyle\frac{1}{2}\xi\textstyle\sum_{\nu} \hat{n}_{\nu}(\hat{n}_{\nu}-\hat{\mathbbm{1}}).
\label{3LSH}
\end{eqnarray}
Here $\hat{a}_{\nu}$, $\hat{a}_{\nu}^{\dag}$, and $\hat{n}_{\nu}$ are bosonic annihilation, creation, and number operators, respectively, for $\nu=\{0,+,-\}$, which represent the 0 and $\pm1$ momentum modes. $E_{\pm}$, from the driving potential in $\hat{H}_\text{B}^\text{Pos}$, couples the 0 and $\pm$ modes, and $\xi$ is the interaction strength with $\xi \equiv U/L=g/[2\pi(N-1)]$. Note that repulsive interactions in the Bose-Hubbard model translate into attractive interactions in the 3LS due to the angular momentum representation\cite{Heimsoth2012}.

\par We can now test the eigenvalue statistics of $\hat{H}_{\text{3LS}}$, which are the quasi-energies of the system. In contrast to the statistics of $\hat{H}_\text{B}^\text{Pos}$, level repulsion becomes manifest in this truncated description with varying amounts depending on the particle interaction strength (Fig.~\ref{levelspacings}). In order to characterize the quantum chaotic nature of the level statistics we fit the eigenvalue spacings of $\hat{H}_{\text{3LS}}$ for $N$ from 10 to 120 with the Brody distribution, $P(s)=b(\eta+1)s^{\eta}\text{exp}(-bs^{\eta+1})$ where $b\equiv\Gamma[(\eta+2)/(\eta+1)]^{\eta+1}$. Here we fit the control parameter $\eta$ which interpolates between regular and quantum chaotic level statistics, that is, $\eta=0$ indicates Poisson statistics and $\eta=1$ is Wigner-Dyson, or quantum chaotic statistics in RMT~\cite{Brody1973}.

\par Figure~\ref{levelspacings}(c) shows the Brody parameter$\eta$ as a function of interaction for particle numbers $N=20$, 50, and 100. Only one dominant trend is observed in the level statistics even though there are, in the semi-classical limit, distinct pockets of chaos on the interval of interaction strengths considered. From this trend we extract the first turning point ($g_s$), maximum ($g_m$), and second turning point ($g_e$), in order to quantitatively characterize the chaotic regime of the system. For this analysis we use a phenomenological modified Lorentzian fit of the form $\eta(g)= A\text{ }\mathrm{exp}[-bg]/(1+c(g+d)^4)$ and ignore the first 15 $\eta$ values since they show fluctuations while being Rabi dynamics. In analogy with the analysis of critical exponents in quantum phase transitions, we also perform a fit of the scaling of these interaction strengths as a function of $N$, $g_{\{s,m,e\}}=A\text{ }N^{B}+C$, and extract their asymptotic values given by the fit parameter $C$. This yields the asymptotic values of $g_s=0.068\pm0.003$, $g_m=0.132\pm0.008$, and $g_e=0.212\pm0.01$, thereby clearly delineating the chaotic regime as shown in Fig~\ref{levelspacings}(d).

\par From the two cases presented in Fig.~\ref{levelspacings} it is clear that the application of RMT as an identifier of quantum chaos has certain limitations. This manifests in our quantum ratchet due to the inclusion of too many non-relevant single particle momentum modes in $\hat{H}_\text{B}^\text{Pos}$ Eq. \eqref{Bose-Hubbard}. It is then clear that for systems in which there is no \textit{a priori} method to anticipate which single particle modes can be truncated RMT is highly limited. Figure ~\ref{levelspacings}(a) explicitly shows failure of RMT for the un-truncated Bose-Hubbard model of our quantum ratchet, while its validity is recovered in the truncated 3LS model, shown in Fig. \ref{levelspacings}(b).

\par In contrast, the proposed dynamical many-body measures of entanglement, condensate depletion, and basis occupation are independent of truncation. However, since the 3LS only includes three momentum modes and has a reduced dimension $\mathcal{D}_\text{3LS}=(N+1)(N+2)/2$ compared to the Bose-Hubbard case $\mathcal{D}_\text{BH} = {{N+L-1}\choose{L-1}}$, we introduce the momentum space Bose-Hubbard model. Considering the momentum space Bose-Hubbard model acquired by the standard discrete Fourier transformation $\hat{b}_i=L^{-1/2}\sum_{k=-\lfloor L/2\rfloor}^{\lfloor L/2\rfloor-1} \hat{a}_k\mathrm{exp}[\mathrm{i}k\theta_i]$ where $\lfloor\cdot\rfloor$ is the floor function, allows us to directly compare the many-body measures regardless of truncation. The new Hamiltonian is then
\begin{eqnarray}
\nonumber \hat{H}_\text{B}^\text{Mom}=&-&2\textstyle\sum_{k=-\lfloor L/2\rfloor}^{\lfloor L/2\rfloor-1} \hat{a}_k^\dag \hat{a}_k\cos(2\pi k/L) \\
\nonumber&-&\frac{U}{2L}\textstyle\sum_{k_j=-\lfloor L/2\rfloor}^{\lfloor L/2\rfloor-1} \hat{a}_{k_1}^\dag\hat{a}_{k_2}^\dag \hat{a}_{k_3} \hat{a}_{k_4} \delta_{k_1+k_2,k_3+k_4} \\
\nonumber&+&\frac{1}{2}\textstyle\sum_{k=-\lfloor L/2\rfloor}^{\lfloor L/2\rfloor-1} [(E_+\mathrm{e}^{\mathrm{i\omega t}}+E_-\mathrm{e}^{-\mathrm{i\omega t}})\hat{a}_k^\dag\hat{a}_{k+\kappa}\\
&+&\textstyle(E_+\mathrm{e}^{-\mathrm{i\omega t}}+E_-\mathrm{e}^{\mathrm{i\omega t}})\hat{a}_{k+\kappa}^\dag\hat{a}_{k}],
\label{BHK}
\end{eqnarray}
where $\hat{a}_k$ and $\hat{a}^\dag_k$ are bosonic annihilation and creation operators in the state with wave number $2\pi k/L$, respectively. Here we note that the $k=0$ and $\pm1$ states are similar to those from the 3LS model but the time dependence remains in the Hamiltonian and is not absorbed into the nonlinear Floquet states as it was in Eq.~\eqref{3LSH}.

\par The dynamics for the remainder of this study were computed using exact diagonalization for every representation (truncation) of the quantum ratchet for a lattice length of $L=10$ and number of particles $N=5$ with the initial state being the ground state of the un-driven Bose-Hubbard. We also perform scaling in the position Bose-Hubbard model Eq.~\eqref{Bose-Hubbard} using TEBD.

\par We apply well established many-body measures: the von Neumann entropy of entanglement $S$, the condensate depletion $D$, and the Inverse participation ratio $P^{-1}$. $S\equiv -\text{Tr}(\rho_A \log\rho_A)$ measures correlations between a subsystem $A$ and the remainder of the system, where $\rho_A$ is the reduced density matrix of a subsystem $A$. For $\hat{H}_\text{B}^\text{Pos}$, there is no symmetry pointing to a particular cut of the system, thus we take $A$ to be half of the ring. Since current reversals are one of the main features of quantum ratchet in its chaotic regime, we take $A$ as the modes with $k\geq0$ for $\hat{H}_\text{B}^\text{Mom}$ and $\hat{H}_\text{3LS}$. The depletion, $D\equiv 1-\lambda_1 /N$ with $\lambda_1$ the first eigenvalue of the single particle density matrix $\langle\hat{a}^\dag_i\hat{a}_j\rangle$, measures the amount of the original condensate still remaining in one single particle mode. Finally, $P^{-1}\equiv (\sum_ip_i^2)^{-1}$, where $p_i$ is the probability of being in the $i$th basis state, and measures the spreading of the state over the many-body Hilbert space.

\par
\begin{figure}
\centering
\includegraphics[width=\linewidth]{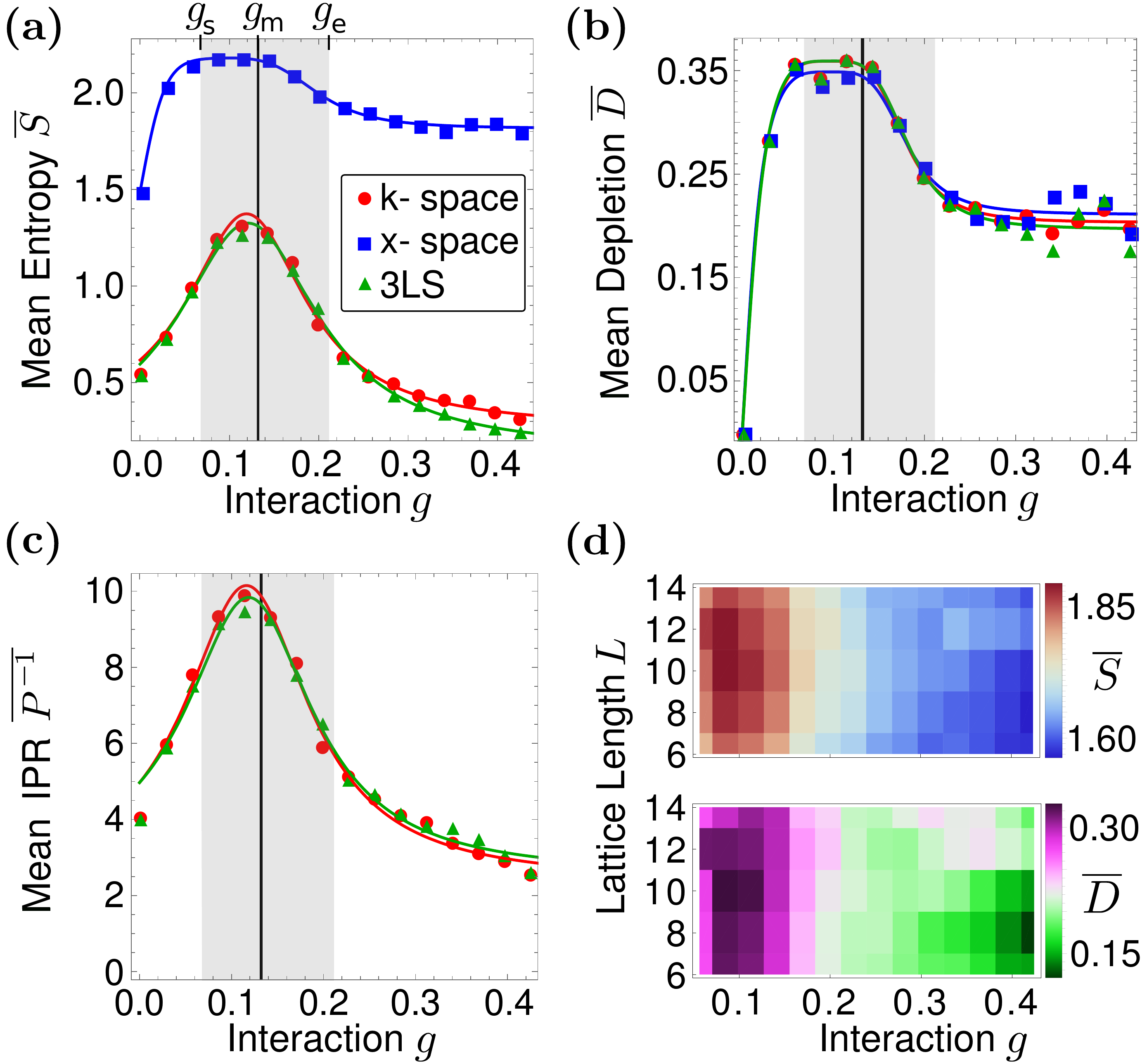}
\caption{\label{means}\emph{Quantum Many-body Measures}. (a)-(c) The chaotic regime is indicated by vertical shading spanning from $g_s$ to $g_e$ with the solid black line giving the maximally chaotic point $g_m$, according to the level statistics in the large $N$ limit. (a) Entropy and (b) depletion show characteristic increases in the chaotic regime in $x$-space, $k$-space, and the truncated Floquet space representations of the ratchet (shared key in (a)). (c) Inverse participation ratio shows a characteristic increase in the chaotic regime only in $k$-space and the truncated Floquet space (shared key in (a)). Each measure allows a fit (solid lines) which can be used to extract the onset of chaos, maximal chaos, and end point of chaos. $(d)$ Entropy and depletion exhibit similar trends regardless of lattice length in $x$-space. The lack of increasing entropy in indicates the system can be simulated using MPS methods.}
\end{figure}

 For each of the above measures we take the time average over 10$T_R$ for 15 values of the interaction strength spanning all three dynamical regimes. We then apply fitting functions to each measure and extract $g_s$, $g_m$, and $g_e$. For the von Neumann entropy we use the fitting functions $\overline{S}(g)=[\tanh(A g)+B][C/(1+D(g+E)^4)+F]$ for $\hat{H}_\text{B}^\text{Pos}$ and $\overline{S}(g)=A/(1+B(g+C)^2)+D]$ for $\hat{H}_\text{B}^\text{Mom}$ and $\hat{H}_\text{3LS}$, see Fig.~\ref{means}(a). For the depletion we use a fitting function $\overline{D}(g)=[\tanh(A g)][B/(1+C(g+D)^4)+E]$ for each model, see Fig.~\ref{means}(b). For the Inverse participation ratio, only $\hat{H}_\text{B}^\text{Mom}$ and $\hat{H}_\text{3LS}$ showed trending with the dynamical regimes, thus these were fit with the function $\overline{P^{-1}}(g)=A/(1+B(g+C)^2)+D]$. The values of the critical interaction strengths $g_s$, $g_m$, and $g_e$ from each measure can be seen in Tab.~\ref{identifiers}.

\par Here we perform a scaling study of the von Neumann entropy and depletion in TEBD for $L=6$ to 14. Figure~\ref{means}(d) shows density plots of the mean entropy (top) and depletion (bottom). The trends are generally the same and can also be fit for the interactions strengths $g_s$, $g_m$, and $g_e$ shown in Fig.~\ref{levelspacings}(d). Figure~\ref{means}(d) clearly shows that the system can be taken to longer lattices without growth in entanglement. This implies that the system can be effectively simulated with matrix product state methods, though full local dimension is required for convergence, and likely to be an issue for other driven and/or highly oscillatory dynamics~\cite{AlcalaDiegoA.GlickJosephA.2016}.

\par Table~\ref{identifiers} summarizes the critical interaction strengths that can be extracted by fitting the many-body measures from exact diagonalization in Fig.~\ref{means}(a)-(c). It is important to note that even though the level statistics of $\hat{H}_\text{3LS}$ predict quantum chaos, the case of $N=5$ for the dynamical measures would be far too small in the 3LS to give any statistical measure of chaos.

\begin{table}
\begin{center}
\begin{tabular}{ c | c | c | c }
Measure &  $g_{s}$ & $g_{m}$ & $g_{e}$ \\ \hline
Level Statistics $N\rightarrow\infty$ & $0.068^{\pm0.003}$ & $0.132^{\pm0.008}$ &  $0.212^{\pm0.01}$\\
Entropy Position & -- & $0.100^{+0.022}_{-0.018}$ & $0.185^{+0.049}_{-0.031}$ \\
Entropy Momentum & $0.072^{+0.005}_{-0.05}$ & $0.119^{+0.002}_{-0.002}$ & $0.166^{+0.005}_{-0.005}$ \\
Entropy Floquet & $0.066^{+0.004}_{-0.004}$ & $0.121^{+0.002}_{-0.002}$ & $0.176^{+0.004}_{-0.004}$ \\
Depletion Position & -- & $0.101^{+0.053}_{-0.021}$ & $0.172^{+0.072}_{-0.029}$ \\
Depletion Momentum & -- & $0.101^{+0.017}_{-0.011}$ &$ 0.169^{+0.020}_{-0.015}$ \\
Depletion Floquet & -- & $0.100^{+0.049}_{-0.020}$ & $0.170^{+0.063}_{-0.027}$ \\
IPR Position & -- & -- & -- \\
IPR Momentum & $0.069^{+0.006}_{-0.007}$ & $0.116^{+0.003}_{-0.003}$ & $0.164^{+0.007}_{-0.006}$ \\
IPR Floquet & $0.070^{+0.007}_{-0.008}$  & $0.119^{+0.003}_{-0.003}$ & $0.167^{+0.008}_{-0.007}$  \\
\end{tabular}
\caption{\label{identifiers}\emph{Tools for Quantum Chaos Identification.} Each measure clearly identifies the maximally quantum chaotic interaction strength $g_m$ and the end of the chaotic regime $g_e$. Only in momentum and Floquet spaces is the onset of the chaotic regime $g_s$ obtained via entropy and inverse participation ratio. Cells with ``--'' indicate a measure fails to predict its corresponding interaction parameter.}
\end{center}
\end{table}

\par In conclusion, we have identified measures that reliably quantify quantum many-body chaos: entanglement, condensate depletion, and inverse participation ratio. Entanglement and inverse participation ratio capture the onset, maximum, and end of chaos in both momentum and the truncated Floquet pictures, while depletion is a universal measure of quantum chaos in all three representations of the quantum ratchet. These measures are especially important for very large systems where obtaining the full spectrum required for random matrix theory (RMT) is computationally inaccessible. Thus, they show potential for identifying quantum chaotic dynamics more generally, whereas RMT is limited in small and large systems, or rapidly becomes misleading due to the inclusion of non-relevant single particle modes. Contrary to the conventional association of quantum chaos with high entanglement and lack of localization in Hilbert space, we have shown for the quantum ratchet that quantum many-body chaos is neither highly entangled, nor does it require many elements of the many-body basis.

\begin{acknowledgments}
We thank Diego Alcala, Daniel Jaschke, and Marie Mclain for their useful discussions. This material is based in part upon work supported by the US National Science Foundation under grant numbers PHY-1306638, PHY-1207881, and PHY-1520915, and the US Air Force Office of Scientific Research grant number FA9550-14-1-0287. Lincoln D. Carr thanks Complutense University for hosting his visit to support the work contained in this Letter. The calculations in this work were executed on the high performance computing cluster maintained by the Golden Energy Computing Organization at the Colorado School of Mines.
\end{acknowledgments}


\end{document}